# Different Response of Molecular Aggregation Structure of Styrenic Triblock Copolymer under Cyclic Uniaxial and Biaxial Stretching Modes

Nattanee Dechnarong[1], Kazutaka Kamitani[2], Chao-Hung Cheng[1], Shiori Masuda[1], Shuhei Nozaki[1], Chigusa Nagano[1], Aya Fujimoto[2], Ayumi Hamada[2], Yoshifumi Amamoto[1,2,3], Ken Kojio[*1,2,3], Atsushi Takahara[**1,2,3]

[1]Graduate School of Engineering, [2]Institute for Materials Chemistry and Engineering, [3]International Institute for Carbon-Neutral Energy Research (WPI-I[2]CNER)
Kyushu University, 744 Motooka, Nishi-ku, Fukuoka, 819-0395, Japan
e-mail: [*]kojio@cstf.kyushu-u.ac.jp, [**]takahara@cstf.kyushu-u.ac.jp

**Abstract**

Mechanical stretching behavior of poly(styrene-*b*-ethylene-*co*-butylene-*b*-styrene) (SEBS) triblock copolymer (87 wt% polyethylene-*co*-butylene (PEB) block, 13 wt% polystyrene (PS) block) was investigated by three different stretching and *in situ* small-angle X-ray scattering (SAXS) measurements. Strain energy density function was investigated based on the stress-stretching ratio ($\lambda$) relationship under uniaxial, planar extension, and equi-biaxial stretching modes. As the result, cross-effect of strain represented by second invariants of the deformation tensor ($I_2$) existed and only Ogden model can be used to fit the data. In the cyclic stretch testing, SEBS exhibited smaller hysteresis during cyclic equi-biaxial stretching mode than for uniaxial stretching one. $\lambda$ and stretching ratio obtained from crystal planes by SAXS ($\lambda_{SAXS}$) were compared to investigate relationship between microdomain structure change and macroscopic mechanical property. SAXS measurement revealed that affine deformation occurred in the smaller $\lambda$ region for both uniaxial and equi-biaxial stretching modes and deviation from affine deformation occurred for uniaxial stretching mode at the larger $\lambda$ region. This is because entangled PEB loop chains could work as cross-linking points when films are stretched by equi-biaxial stretching mode.

## 1. Introduction

Elastic property is the ability that materials can return to their original shape and size when external force is removed. For rubbers and elastomers, the elasticity originates from their network structures. Two requirements of the elastic property are retractive force and entropic nature of the rubber network. During stretching, rubber chains changed their conformation from gauche rich to trans rich state, leading to a decrement of entropy. This change in entropy produces retractive force, which governs elastic property of the rubbers [1]. When the rubbers and elastomers are subjected to cyclic deformation, hysteresis appears at large strain when unloading. The phenomenon is well-known as the Mullins effect [2-6]. Several physical interpretations have been proposed to explain the Mullins effect, including bond rupture in the network structure, filler rupture, molecular slippage, and disentanglement. Breaking of the interaction between rubber and fillers are also a possible causes of the Mullins effect [7]. In styrenic triblock copolymers which is a type of elastomers, pulling out of the rubbery midblock chains from its cross-linking points might be a cause of Mullins effect.

Most studies of Mullins effect were discussed based on cyclic uniaxial stretching [8]. Nevertheless, discussion based on only uniaxial elongation mode is insufficient for explaining the deformation behavior of a material [9, 10]. An observation of the Mullins effect during various type of deformation is important to understand the softening behavior of the sample [6, 11-14]. Cyclic biaxial stretching is one of an efficient elongation mode to understand the deformation mechanism of samples, as information from this mode can yield information on many aspects of nonlinear stress-strain relationship. Urayama *et al.* investigated the Mullins effect for silica filled-reinforced styrene butadiene elastomer comparing under three types of extension, *i.e.*, uniaxial, planar and equi-biaxial extension [14]. Energy dissipation and residual strain increased in the order of equi-biaxial, planar and uniaxial extension with increasing volume fraction of fillers. They concluded that filler network and interaction between the fillers and the rubber matrix is a major factor to govern the dissipation.

Since physical cross-linking points are employed in styrenic triblock copolymers, these systems are environmental-friendly elastomers. Poly(styrene-*b*-ethylene-*co*-butylene-*b*-styrene) (SEBS) consists of rubbery poly(ethylene-*co*-butylene) (PEB) midblocks and glassy PS end blocks. Fractions of PS and PEB regulate the properties of SEBS. PS domains form physical cross-linking points and have a role to keep a network structure, while the PEB chains give large stretchable property [15, 16]. There are two possibilities for midblock structures: bridge and loop chains [17, 18]. Bridge chains are segments that connect different PS domains, whereas loop chains are segments in the same PS domain. As fraction and conformation of bridge and loop chains are closely related to tensile properties, they have to be taken into consideration [17, 19]. We reported changes in microphase-separated structure of SEBS using small-angle X-ray scattering (SAXS) [20]. SAXS can clarify the change in a microdomain structure in detail [21-30]. In this work, effect of deformation mode and cyclic deformation on mechanical stretching behavior of SEBS was investigated using SAXS measurement.

## 2. Experiments
2.1 Sample preparation

A SEBS sample with 13 wt% PS was provided by Asahi Kasei Chemical Co., Ltd, Japan. The chemical structure of SEBS is illustrated in Figure S1 in Supporting Information. The characteristics of the SEBS sample are specified in Table 1. SEBS pellets were dissolved in toluene and precipitated in methanol to remove impurities. Pieces of the sample were dried at room temperature and then under high-pressure vacuum until the weight stabilized. Purified samples were pressed into films at 150°C without the application of pressure for 5 min, and then a pressure of 10 MPa was continuously applied for 1 min. Films were obtained with a thickness of 300-400 μm. To obtain the periodic structure of microphase separation, samples were annealed at 170°C for 7 days in vacuo. The annealed films were cooled to room temperature once annealing was finished. Then, specimens were cut into a rectangular shape with dimensions of 5×30 mm$^2$ for the cyclic uniaxial test and 20×20 mm$^2$ for the cyclic equi-biaxial test.

**Table 1.** Characteristics of SEBS

| Sample | PS content[a] (wt %) | $\bar{M}_n$[b] | $\bar{M}_w$[b] | PDI[b] | $T_g$ of PEB[c] (°C) | $T_g$ of PS[c] (°C) |
|---|---|---|---|---|---|---|
| SEBS-13 | 13 | 112,000 | 163,000 | 1.46 | -40 | 52 |

[a]PS content was determined from $^1$H-NMR. [b]Molecular weight information was determined from GPC. [c]Glass transition temperature ($T_g$) was determined from DSC.

2.2 *In situ* SAXS measurement during cyclic stretching

For the cyclic uniaxial and equi-biaxial tests, each sample was clamped between the chuck grips of custom-built tensile testing machines (uniaxial machine: DIP Co.; biaxial machine: JUNKEN MEDICAL Co., Ltd), which were enabled for *in situ* measurement. Film samples were stretched with loading-unloading cycles with increasing maximum stretching ratio ($\lambda_m$) of each cycle under uniaxial and equi-biaxial elongation at room temperature at 1 mm s$^{-1}$. For cyclic uniaxial stretching, the sample was stretched for five cycles to $\lambda_m$ = 2.3, 3.6, 4.8, 6.1, and 7.4, respectively. For cyclic equi-biaxial stretching, the sample was stretched for four cycles to $\lambda_m$ = 2.3, 3.0, 3.6, and 4.2, respectively. The synchrotron SAXS measurement was conducted with the BL40XU and BL05XU beamlines in the SPring-8 facility in Japan. The beam size at the samples was 150 × 150 μm$^2$. The wavelength of the X-ray was 0.100 nm, and the sample-to-detector distance was ca. 2 m. The samples were exposed to the X-ray beam for 0.5-1 s at 25°C. SAXS patterns were taken with the beam perpendicular to the film surface (through view). 2D-scattering patterns of SAXS were obtained from a PILATUS 100k detector (DECTRIS Ltd) with a total pixel size of 172×172 μm$^2$. Data were converted from 2D patterns to 1D profiles by integrating with FIT2D (ver. 16.041, Andy Hammersley/ESRF, Grenoble, France).

## 3. Results and discussion
### 3.1 Strain energy density function analysis

The stress-stretching ratio ($\lambda$) relation is governed by strain energy density function ($W$). Several models of strain energy function were considered to investigate the dominant factors in mechanical properties of samples [10, 31-34]. In this study, $W$ of four models was investigated based on the stress-$\lambda$ relationship of equi-biaxial stretching (EB) and planar extension (PE) of SEBS. The films were stretched in X-axis ($\lambda_x$) with maintaining the initial dimension in Y-axis ($\lambda_y$ = 1) during the planar extension mode. The classical neo-Hookean model was initially explored as it is the simplest model among other existed constitutive models. This model describes ideal rubber networks with infinite extensibility and without structural defects. $\lambda_x \lambda_y \lambda_z = 1$ is employed due to the incompressibility. Since SEBS showed strain hardening at high deformation region which is a characteristic of non-Gaussian statistics, Gent, extended Gent [32] and Ogden models [10, 34] were applied to investigate the relationship of stress and $\lambda$ of SEBS as they have been used to describe the non-linear elasticity of various elastomeric materials. Equations of $W$ and $\sigma$ of neo-Hookean, Gent and extended Gent models are shown in the Supporting Information. $W$ of the Ogden model is expressed as

$$W = \sum_n \frac{\mu_n}{\alpha_n} \left( \lambda_x^{\alpha_n} + \lambda_y^{\alpha_n} + \lambda_z^{\alpha_n} - 3 \right) \qquad (1)$$

The additional terms of $\mu$ and $\alpha$ were introduced to achieve a good fitting to the experimental data. The stress-$\lambda$ relationships of Ogden model are shown in following equations:

$$\sigma_{EB} = \frac{1}{\lambda_x} \sum_n \mu_n \left( \lambda_x^{\alpha_n} - \lambda_x^{-2\alpha_n} \right) \qquad (2)$$

$$\sigma_{PE-X} = \frac{1}{\lambda_x} \sum_n \mu_n \left( \lambda_x^{\alpha_n} - \lambda_x^{-\alpha_n} \right) \qquad (3)$$

$$\sigma_{PE-Y} = \sum_n \mu_n \left( 1 - \lambda_x^{-\alpha_n} \right) \qquad (4)$$

$\sigma_{EB}$, $\sigma_{PE-X}$ and $\sigma_{PE-Y}$ are defined as stress of equi-biaxial stretching, planar extension in X- and Y-axes, respectively.

Figure 1 (a) and (b) illustrates relationships between $\lambda$ and stress of SEBS during equi-biaxial stretching and planar extension, and the data at small $\lambda$ region, respectively. The initial moduli of equi-biaxial stretching, planar extension in X-axis, and Y-axis were consistent with the infinitesimal linear elasticity theory as these moduli were approximately $6G$, $4G$, and $2G$, respectively [35]. Plots of model analysis are shown in Figure 2. It was found that the prediction of neo-Hookean model did not fit to the experimental data, as shown in Figure 2 (a). This is because the tensile properties of SEBS were not consistent with the assumption of the neo-Hookean model. With the consideration of finite extensibility effect, Gent, extended Gent and Ogden models were introduced to determine the tensile properties of SEBS. From Figure 2 (b), Gent model did not fit well with the experimental data of SEBS, indicating that additional factor should be considered. The value of stress ratio ($\sigma_y/\sigma_x$) of planar extension test with various $\lambda_x$ was obtained in order to investigate the coupling between X- and Y-axis, as shown in Figure 2 (e). It was found that $\sigma_y/\sigma_x$ lies above the prediction of neo-Hookean and Gent model (dashed line), suggesting that the cross-effect of strains in different directions affects the deformation of SEBS. Therefore, extended Gent model with consideration of the cross-effect of strain represented by second invariants of the deformation tensor ($I_2$) was applied. However, it was found that extended Gent model did not show the good agreement with the experimental data of SEBS, as shown in Figure 2 (c). The advantage of Gent and extended Gent model is that the parameters from these models can be interpreted the molecular properties of elastomers, however, they are unable to predict the properties in whole range of strain of SEBS. Therefore, Ogden model was employed in this study. It was found that experimental data of SEBS during equi-biaxial and planar extension testes was satisfactorily described by the Ogden model using two terms (n=2) of the single set parameter, as shown in Figure 2 (d).

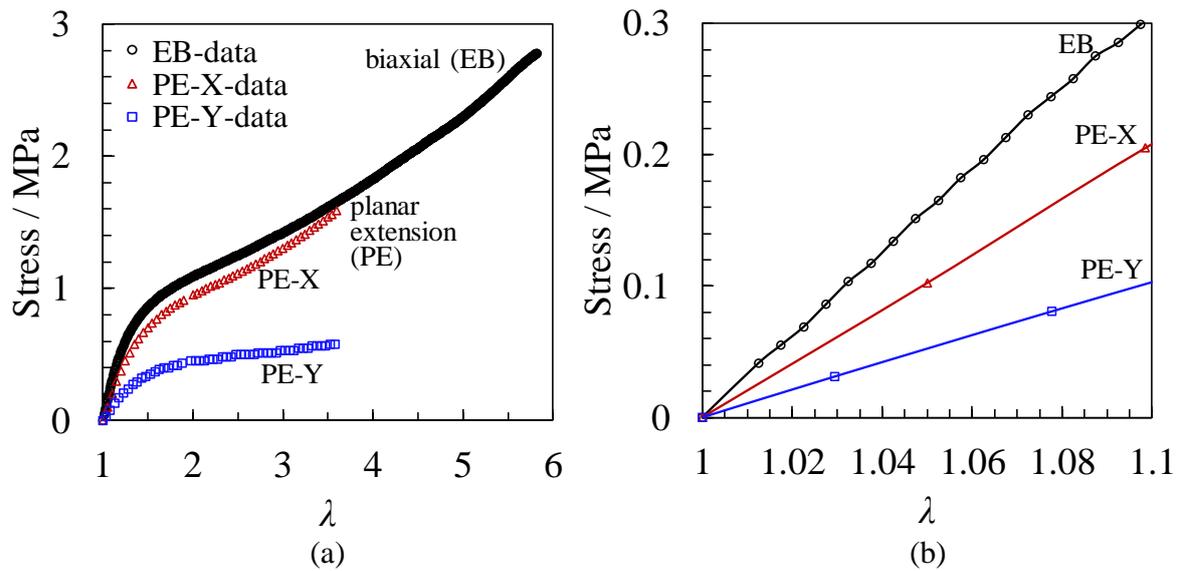

**Figure 1.** (a) Relationships between $\lambda$ and stress of SEBS during equi-biaxial stretching and planar extension (experimental data). (b) Data of (a) at small deformation region.

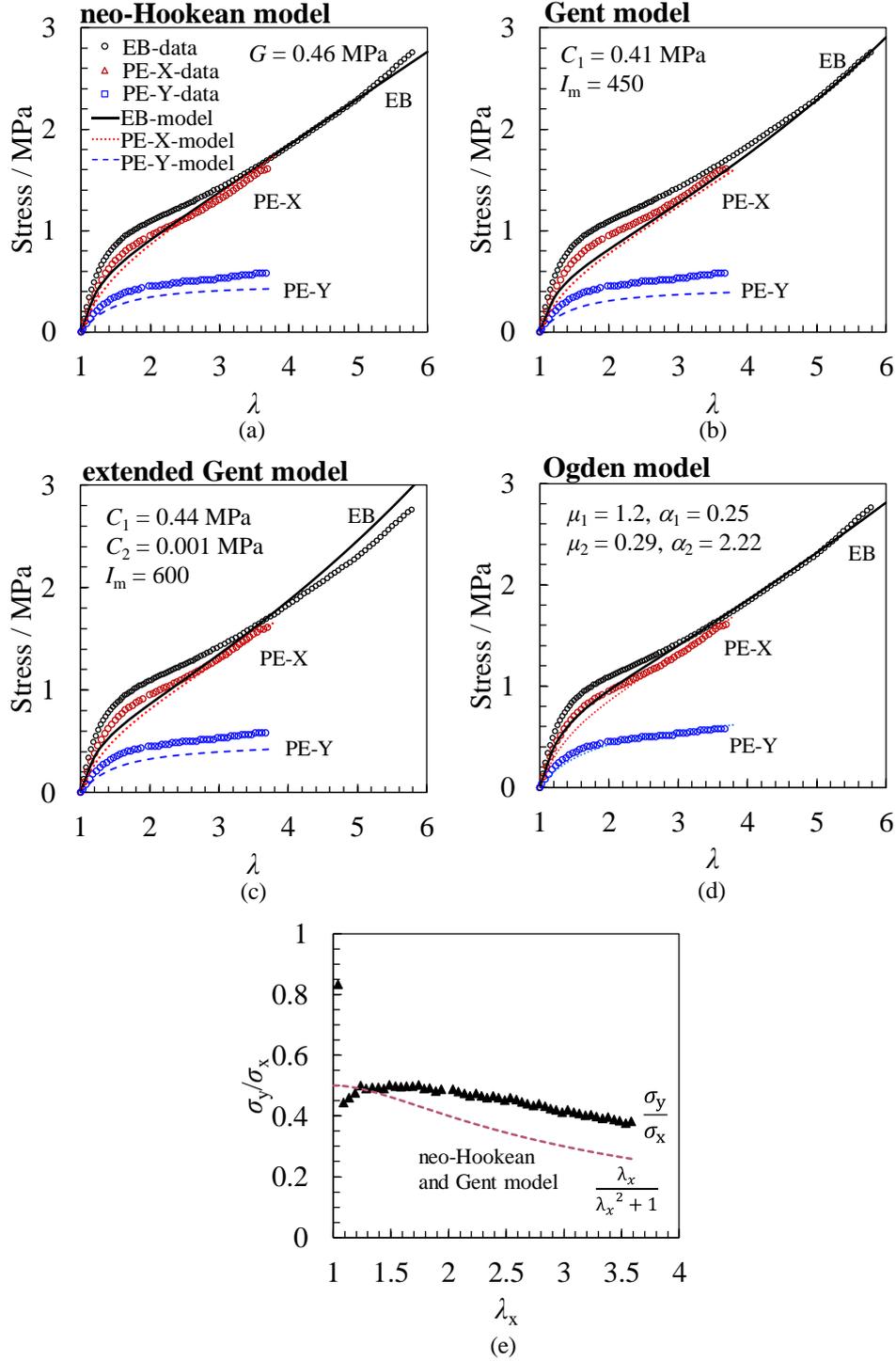

**Figure 2.** Relationships between real $\lambda$ and stress of SEBS during equi-biaxial stretching and planar extension. Line curves correspond to the prediction of (a) neo-Hookean model, (b) Gent model, (c) extended Gent model and (d) Ogden model. (e) Stress ratio ($\sigma_y/\sigma_x$) as a function of $\lambda_x$ in planar extension of SEBS. The dashed line is the prediction of the neo-Hookean and Gent model.

3.2 Stress-$\lambda$ relationship of SEBS during cyclic stretching

  To investigate mechanical stretching behavior of SEBS, films were stretched with loading-unloading cycles by various $\lambda_m$s under uniaxial and equi-biaxial elongation at 25°C. Figure 3 (a) and (b) shows stress-$\lambda$ relationship of cyclic uniaxial and equi-biaxial stretchings at 1 mm s$^{-1}$ and 25°C, respectively. Stress during loading and unloading process exhibited different magnitude for cyclic uniaxial stretching as shown in Figure 3(a). This difference became larger with increasing $\lambda_m$. Permanent set also increased with $\lambda_m$. On the contrary, stress during loading and unloading process exhibited similar value for cyclic equi-biaxial stretching and smaller permanent set, as shown in Figure 3 (b). In other word, smaller hysteresis was observed in the results of cyclic equi-biaxial stretching than that of cyclic uniaxial stretching mode.

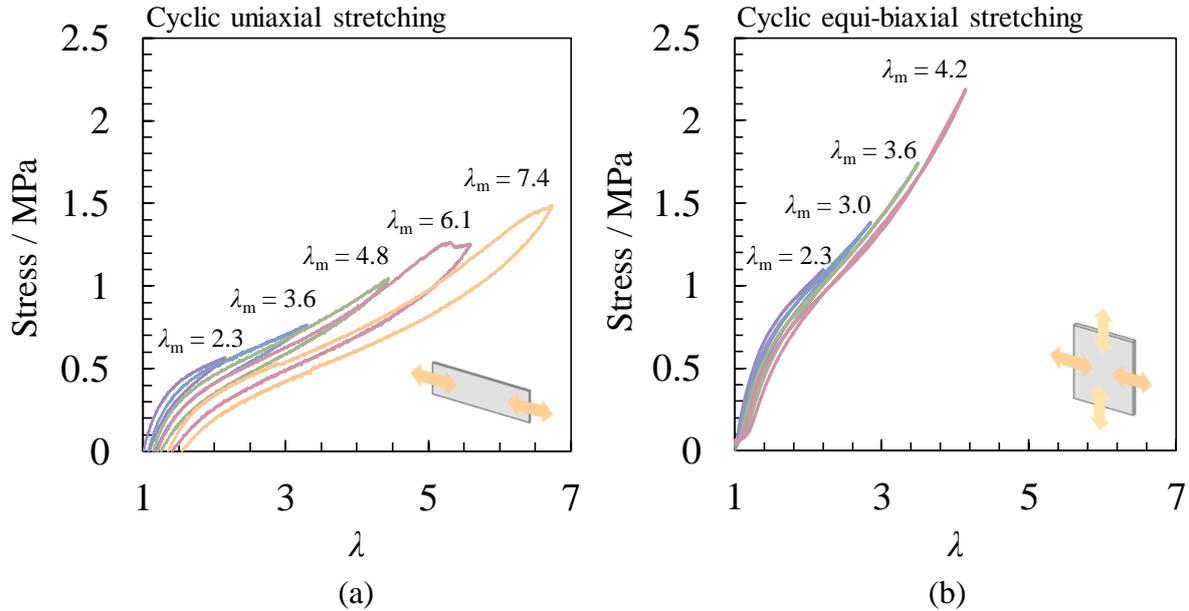

**Figure 3.** Stress-$\lambda$ relationship of SEBS under (a) cyclic uniaxial stretching and (b) cyclic equi-biaxial stretching at 1 mm s$^{-1}$ and 25°C.

Quantitative analysis was carried out to discuss the hysteresis of two deformation modes based on the calculation of dissipation energy ($D$). Figure 4 shows relationship between dissipation factor ($\Delta$) and $\lambda_m$ of each cycle of cyclic uniaxial and equi-biaxial elongation. The $\Delta$ was calculated from a ratio of $D$ to stored elastic energy in the virgin loading process ($W_0$). The $\Delta$ can be used to compare $D$ of different deformation modes and $\lambda$ with consideration to $W_0$ [14]. $D$ is energy consumption of the rubber chains during the cyclic deformation. High $D$ indicates high hysteresis. The equations are as follow:

$$\Delta = \frac{D}{W_0} \qquad (5)$$
$$D = W_{load} - W_{unload} \qquad (6)$$

where $W_{load}$ and $W_{unload}$ are stored elastic energy in loading process and released elastic energy in unloading process, respectively. It was found that $\Delta$ obviously increased with increasing $\lambda_m$ of each cycle during cyclic uniaxial stretching. In the case of cyclic equi-biaxial stretching, $\Delta$ did not increase with increasing $\lambda_m$ of each cycle. This difference in the energy dissipation in two modes was suggested to be mainly from the network of PEB chains, which give stretchable property in SEBS. Energy loss in SEBS during the cyclic test could be due to an occurrence of irreversible structure change during deformation, such as pulling out and slipping of PEB chains from the PS domains. Discussion in detail will be made in the later section.

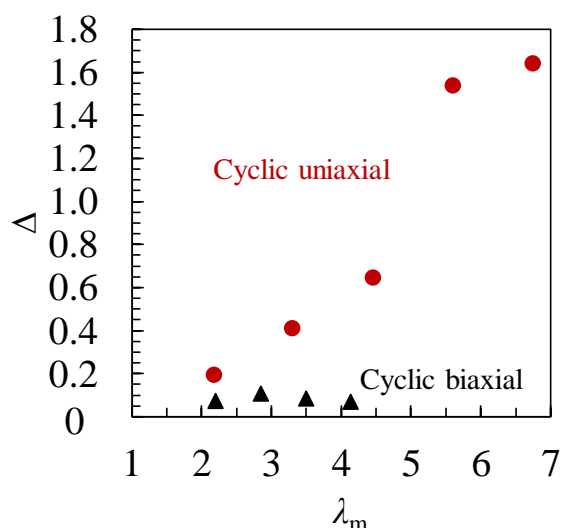

**Figure 4.** Relationship between dissipation factor (Δ) and $\lambda_m$ of each cycle during cyclic uniaxial and equi-biaxial stretching at 1 mm s$^{-1}$.

3.3 Change in microphase-separated structure of SEBS during cyclic stretching

*In situ* SAXS measurement was carried out to investigate an arrangement of PS domains. 2D SAXS patterns and 1D SAXS profiles of the SEBS film before and after annealing are illustrated in Figure S2 in Supporting Information. Three sharp rings were observed in the scattering pattern of the annealed sample. As a ring-like pattern appeared instead of a spot pattern, the size of randomly oriented grains was much smaller than for X-ray beam. By conversion of the 2D pattern to the 1D profile along various directions, the three sharp rings transformed into three diffraction peaks. These crystalline peaks were observed at scattering vectors ($q$) of 0.26, 0.36, and 0.45 nm$^{-1}$. These peak positions correspond to the $q$ relation of 1, √2 and √3, which can be assigned to the (110), (200), and (211) planes of the body-centered cubic (bcc) lattice, respectively [36]. Fringes of intensity in the profile correspond to the form factor, which indicates the shape and size of PS domains. The SEBS showed the characteristic of the formation of spherical PS domains.

As PS domains are the physical cross-linking points in the PEB matrix, a measurement of an arrangement of PS domains using *in situ* SAXS measurement can be used to explain changes in the rubber network. Figures 5 and 6 show 2D SAXS patterns and 1D SAXS profiles of SEBS during cyclic uniaxial and equi-biaxial stretching modes, respectively. Discussion of the cyclic uniaxial stretching mode would be focused on an arrangement of PS domains in stretching direction. Three diffraction peaks shifted to lower $q$ region during loading process and continuously shifted back during unloading process. These shifts referred to a change in plane spacing ($d$) during stretching. Plane spacing of PS domains increased with increasing $\lambda$ during loading process and decreased with decreasing $\lambda$ during unloading process. In the case of cyclic equi-biaxial stretching mode, three diffraction peaks also shifted to lower region during loading process and shifted back during unloading process. However, it was found that the position of the diffraction peaks after finished the unloading process of each cycle slightly shifted to lower $q$ region. This shift of the diffraction peaks from an initial position implies that plane spacing of PS domains did not totally return to its original position, however, the plane spacing slightly increased after finished the unloading process. An increase in the plane spacing implies that residual $\lambda$ slightly occurred in structure of SEBS during cyclic equi-biaxial stretching mode. The occurrence of residual $\lambda$ in cyclic equi-biaxial stretching could be because simultaneous stretching in two directions.

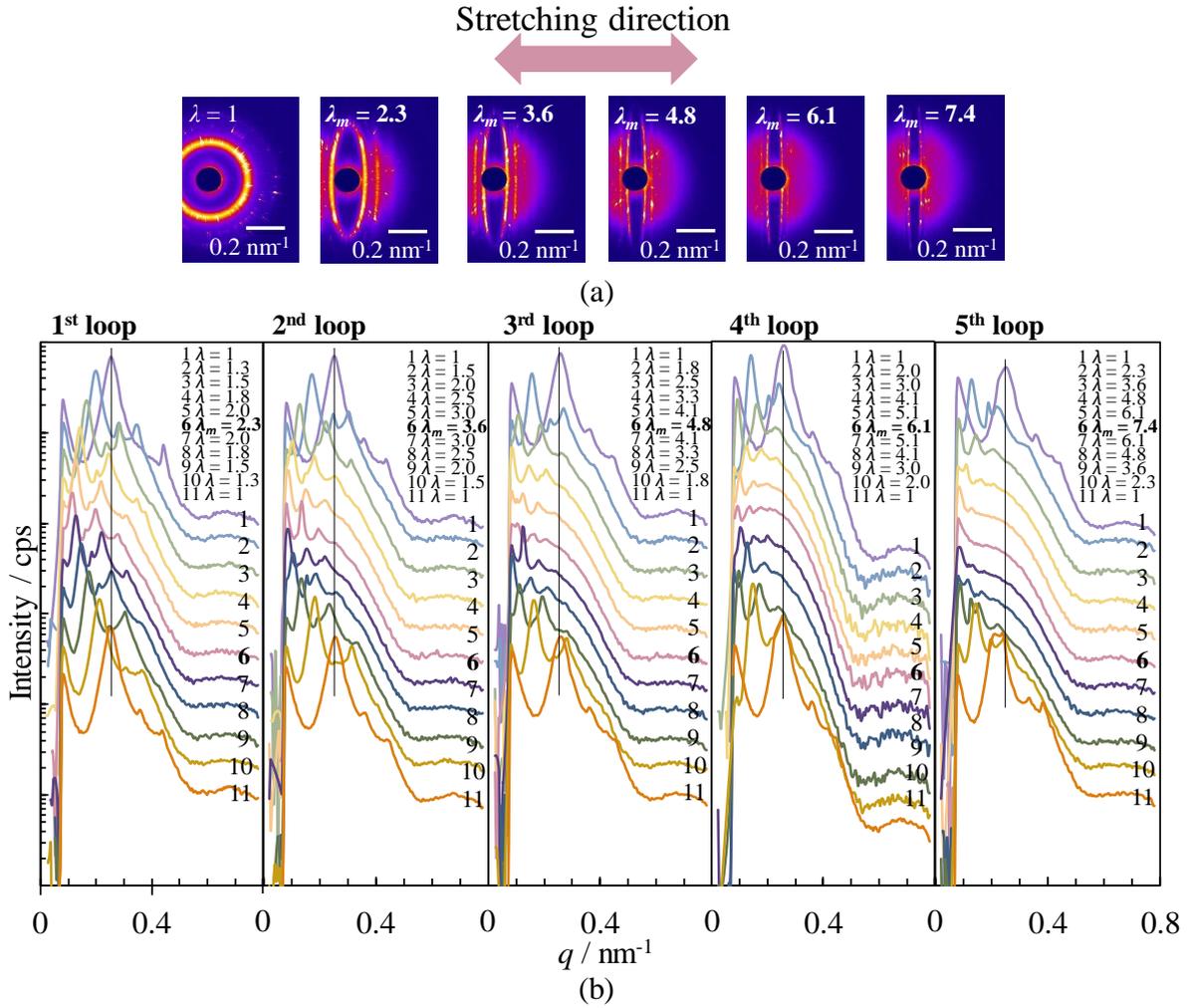

**Figure 5.** (a) 2D SAXS patterns and (b) 1D SAXS profiles of SEBS obtained from (a) in the stretching direction during cyclic uniaxial stretching at 1 mm s$^{-1}$ and 25°C at various $\lambda$s.

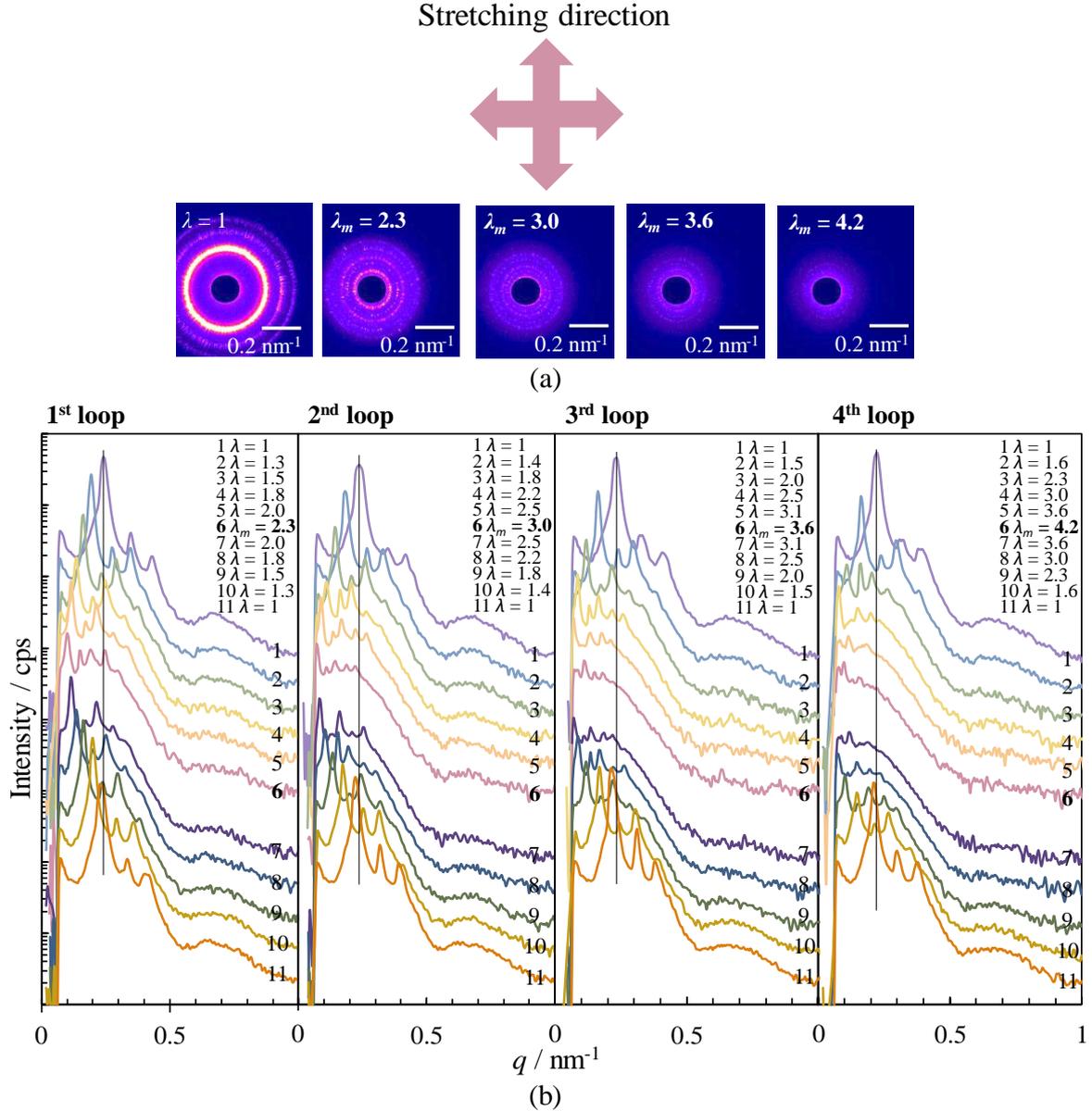

**Figure 6.** (a) 2D SAXS patterns and (b) 1D SAXS profiles of SEBS obtained from (a) in X-axis during cyclic equi-biaxial stretching at 1 mm s$^{-1}$ and 25°C at various $\lambda$s.

Figure 7 (a) shows relationship between $\lambda$ and $\lambda$ of (110) plane spacing obtained from SAXS ($\lambda_{SAXS(110)}$) with various $\lambda_m$. The diagonal lines shown in the Figure 7 (a) correspond to affine deformation. In the virgin loading of cyclic uniaxial stretching ($\lambda_m = 2.3$), SEBS showed affine deformation, indicating the simple arrangement of PS domains. This result is consistent with the results of the elongation at low $\lambda$ of simple stretching. Deviation from the affine deformation tended to occur with increasing $\lambda_m$, indicating that ordering of arrangement of PS domains decreased with increasing the number of elongation cycle. Irreversible structure change, such as pulling out of PEB bridging chains from PS end blocks, breaking of PEB chains, and breaking of PS domains, might occur during cyclic stretching. This irreversibility correlated to deviation of an arrangement of PS domains from the affine deformation. Degree of the deviation increased with increasing $\lambda_m$ of each cycle. The $\lambda_{SAXS}$ of unloading process showed slightly larger deviation than loading process of each cycle. This result correlated to the hysteresis observed from the curves of stress-$\lambda$ relationship.

Figure 7 (b) shows $\lambda$ of plane spacing obtained from SAXS ($\lambda_{SAXS}$) with various orientation angles of crystal plane ($\omega$) during cyclic uniaxial stretching. The deviation was observed in various orientation of the crystal plane in SEBS, as shown in Figure 7 (b). The $\lambda_{SAXS}$ measured at various $\omega$ at the $\lambda$ of 1.51 and 2.28 of the virgin loading showed a good agreement with the theoretical curve of the affine deformation (line curves), which was calculated from equation (7) [37].

$$\lambda_{\text{theory SAXS}} = [\alpha_x^2 cos^2\omega + (1/\alpha_x)sin^2\omega]^{1/2} \qquad (7)$$

where $\alpha_x$ is macroscopic strain and $\omega$ is orientation angle of crystal plane. Additional explanation is included in Figure S3 in the Supporting Information. On the other hand, the unloading and reloading process obviously exhibited the deviation from the theorical affine deformation in various orientation of the crystal plane. At the $\lambda$ of 4.84 of the third cycle, deviation from the affine deformation was obviously observed.

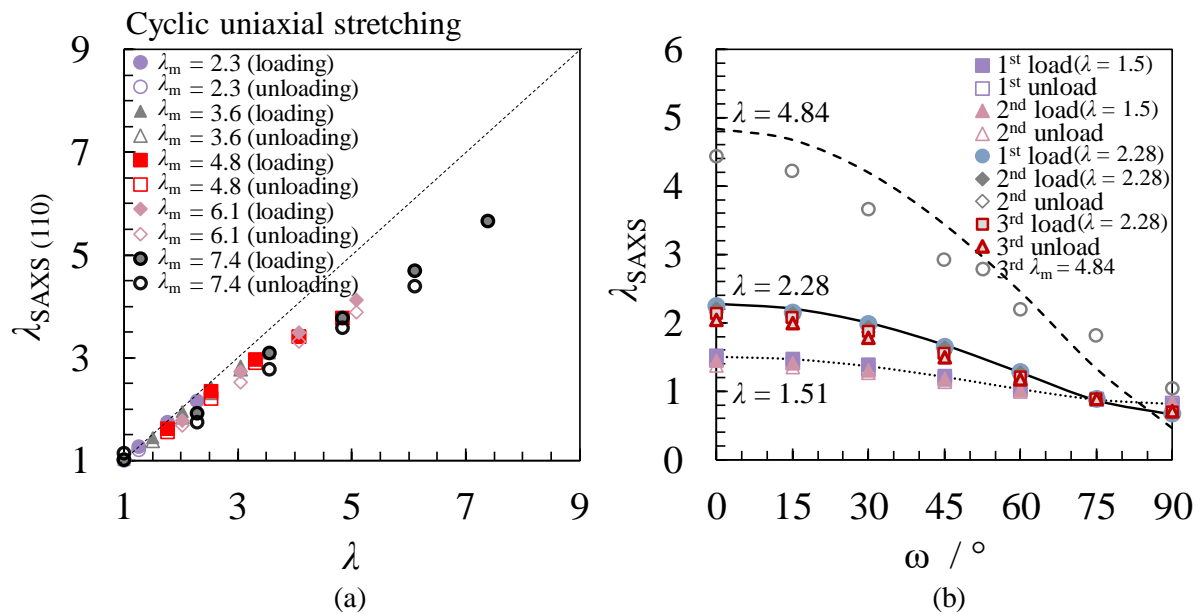

**Figure 7.** Relationship between (a) $\lambda$ of SEBS and $\lambda_{SAXS(110)}$ with various $\lambda_m$s during cyclic uniaxial stretching at 1 mm s$^{-1}$, (b) $\lambda_{SAXS}$ with various $\omega$s during cyclic uniaxial stretching at 1 mm s$^{-1}$. Line plots are the theoretical curves of the affine deformation.

Figure 8 shows relationship between $\lambda$ and $\lambda_{SAXS}$ at $\lambda_m$s of (a) 2.3, (b) 3.0, (c) 3.6 and (d) 4.2 during cyclic equi-biaxial stretching. In the case of cyclic equi-biaxial stretching, SEBS also showed the simple arrangement of PS domains in the virgin loading process. Then, the deviation slightly occurred with increasing $\lambda_m$. However, this deviation during cyclic equi-biaxial stretching was smaller than during cyclic uniaxial stretching mode. These results are correlated to the hysteresis result. In the larger strain region, the crystal plane-dependence on the affine deformation was observed during cyclic equi-biaxial deformation mode. Figure 9 shows schematic illustration of an arrangement of PS domains in (110), (200), (211), (220) and (310) crystal planes of bcc lattice. Crystal planes with small periods, like (220) and (310), tended to show large deviation. Two main bridge PEB chains were assumed as 1$^{st}$ nearest and 2$^{nd}$ nearest bridges (($\sqrt{3}/2$)$a$ and $a$), where $a$ is lattice length of bcc lattice. For the planes of (110), (200), and (211), the direction of bridges is close to perpendicular to each plane, however, that for (310) is close to parallel. Furthermore, the number of the PEB chains between each (310) plane are fewer than for other planes, and there is no PEB chains between PS domains placed in (310) planes. Thus, elastic property of (310) planes might be low in comparison with those of (110), (200), and (211), resulting larger deviation of (310) from affine deformation. On the other hand, this plane-dependence was not observed in cyclic uniaxial stretching mode [20].

Smaller hysteresis and deviation from affine deformation of cyclic equi-biaxial stretching mode can be discussed based on the correlation between types of PEB bridge chains and cross-effect of strain. Various types of PEB bridge chains in the network structure of SEBS were presumed. Each type of bridge chains has different length, defined as $1^{st}$, $2^{nd}$, $3^{rd}$, and $4^{th}$ nearest chains. This difference in chain length might be a reason of the deviation of $\sigma_y/\sigma_x$ from the prediction of neo-Hoookean and Gent models, as shown in Figure 2 (e). Thus, cross-effect strain became dominated. Entanglement effect has been proposed to be one of the origin of the deviation from the classical theories [38]. For SEBS, entanglement effect seemed to occur during equi-biaxial stretching more than during uniaxial stretching, confirming by smaller strain at break of equi-biaxial stretching mode. This entanglement might be induced by topological effect during equi-biaxial stretching. Figure 10 shows schematic illustration of the occurrence of nominal cross-linking points during equi-biaxial stretching. Entangled PEB chains might produce the temporary cross-linking points in addition to the physical cross-linking points, PS domains, bringing about smaller hysteresis during equi-biaxial stretching.

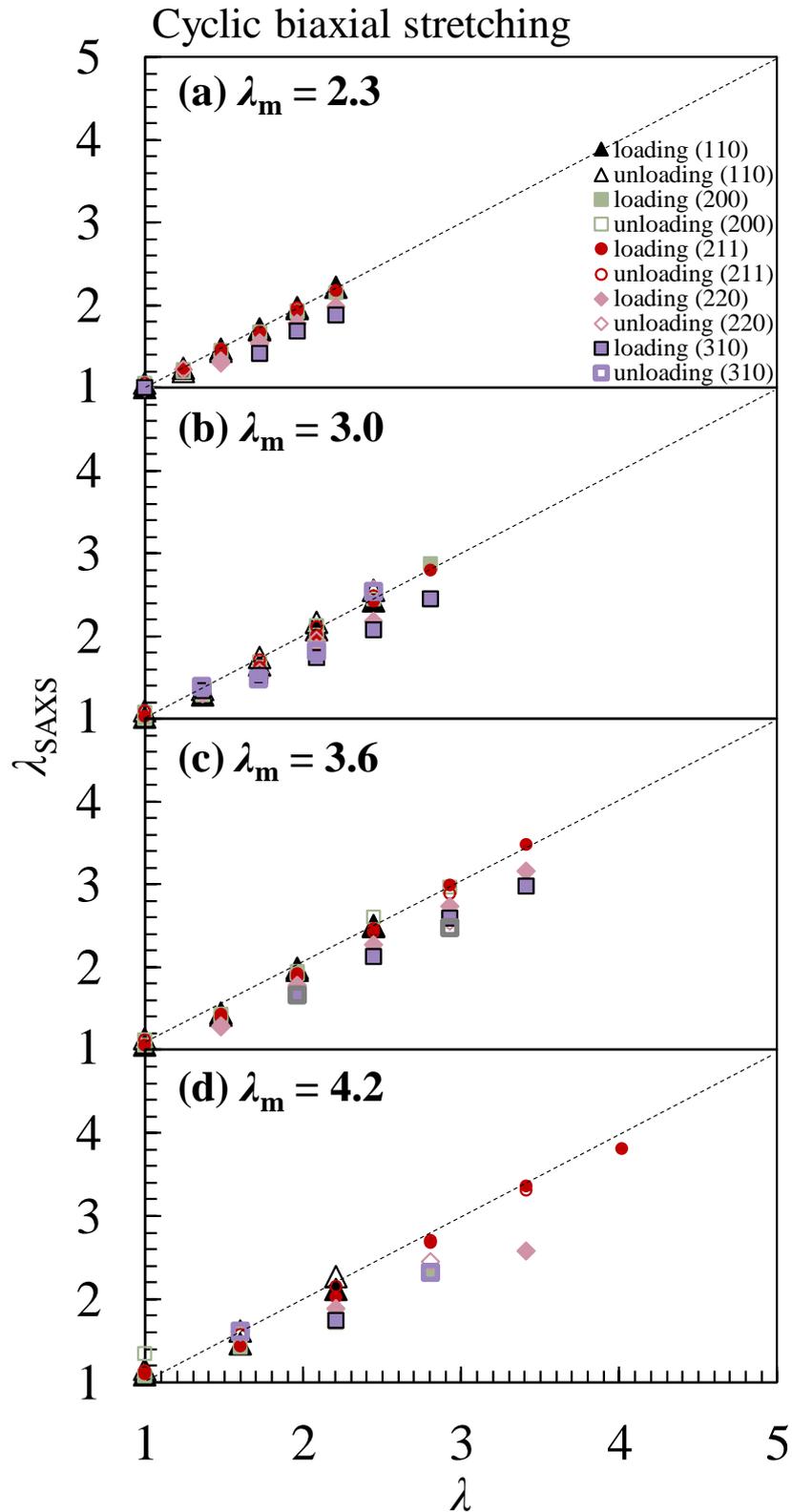

**Figure 8.** Relationship between $\lambda$ and $\lambda_{SAXS}$ with various $\lambda_m$s of (a) 2.3, (b) 3.0, (c) 3.6, and (d) 4.2 of each cycle during cyclic equi-biaxial stretching at 1 mm s$^{-1}$.

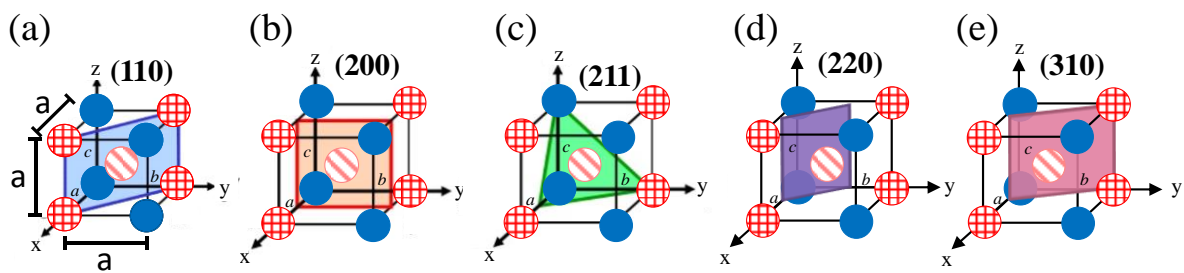

**Figure 9.** Schematic illustration of an arrangement of PS domains in (a) (110), (b) (200), (c) (211), (d) (220) and (e) (310) crystal planes of bcc lattice.

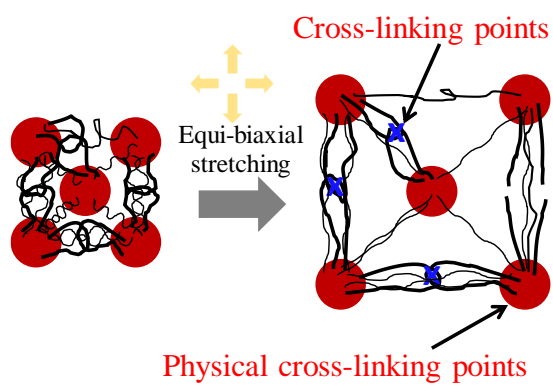

**Figure 10.** Schematic illustration of the occurrence of nominal cross-linking points during equi-biaxial stretching.

## 4. Conclusions

Two terms (n = 2) of the single set parameters from the Ogden model successfully described the relationships of stress and $\lambda$ of SEBS. Hysteresis and deviation from the affine deformation confirmed an occurrence of the irreversible structure change in SEBS. Cyclic equi-biaxial stretching exhibited smaller deviation from the affine deformation than cyclic uniaxial stretching, which correlated to hysteresis. Cross-effect of strains in different directions which has been attributed to the entanglement effect clarified the smaller hysteresis during cyclic equi-biaxial stretching. Cross-linking points might be produced from entangled PEB loop chains when films are stretched by equi-biaxial stretching mode.

**Acknowledgments**

This work was supported by the Impulsing Paradigm Change through Disruptive Technology (ImPACT) Program, JST CREST Grant Number JPMJCR17J4 and JST-Mirai Program Grant Number JPMJMI18A2, Japan. Synchrotron radiation X-ray scattering measurements were performed at BL40XU and BL05XU in the SPring-8 facility with the approval of the Japan Synchrotron Radiation Research Institute (JASRI; Proposal No. 2018B1035, 2019A1015, 2019B1011, 2020A1007). N.D. gratefully thanks Weeradet Sittiphon for his support in mathematical calculation. Financial support for N.D. was provided by the International Graduate Course on Chemistry for Molecular Systems, Ministry of Education, Culture, Sports, Science and Technology (MEXT), Japan.